# Smart Service-Oriented Clustering for Dynamic Slice Configuration

Tarik Taleb[1,2,3], Djamel Eddine Bensalem[1], and Abdelquoddouss Laghrissi[1,2]

[1] Aalto University, Espoo, Finland

[2] University of Oulu, Oulu, Finland

[3] Sejong University, Seoul, Korea

Emails: firstname.lastname@aalto.fi

*Abstract*—The fifth generation (5G) and beyond wireless networks are foreseen to operate in a fully automated manner, in order to fulfill the promise of ultra-short latency, meet the exponentially increasing resource requirements, and offer the quality of experience (QoE) expected from end-users. Among the ingredients involved in such environments, network slicing enables the creation of logical networks tailored to support specific application demands (i.e., service level agreement SLA, quality of service QoS, etc.) on top of physical infrastructure. This creates the need for mechanisms that can collect spatiotemporal information on users' service consumption, and identify meaningful insights and patterns, leveraging machinelearning techniques. In this vein, our paper proposes a framework dubbed "SOCL" for the Service Oriented CLustering, analysis and profiling of users (i.e., humans, sensors, etc.) when consuming enhanced Mobile BroadBand (eMBB) applications, internet of things (IoT) services, and unmanned aerial vehicles services (UAVs). SOCL relies mainly on the realistic network simulation framework "network slice planner" (NSP), and two clustering methods namely K-means and hierarchical clustering. The obtained results showcase interesting features, highlighting the benefit of the proposed framework.

## I. INTRODUCTION

To meet different verticals with diverse service and resource requirements [1], 5G will rely not only on the advancement of radio access network (RAN) (i.e., millimeter waves, new radio frequencies, beam-forming, massive MIMO, etc.), but also on key enabler technologies for network softwarization such as software defined network (SDN) [2], [3] and network function virtualization (NFV) [4]. Leveraging these technologies, 5G is expected to achieve high-bandwidth, ultra-low latency, and high-density connections, thus, enabling multiple use cases that were not possible in the previous network generations.

In this vein, the 5G network is envisioned to support a wide range of distinct services for different types of equipment varying from regular terminals such as smartphones and tablets to a broader set of devices such as UAVs, autonomous cars and IoT sensors. The International Telecommunication Union (ITU) and Fifth Generation Public Private Partnership (5GPPP) encapsulate the use-cases into three main categories [5]: eMBB, massive machine-type communications (MTC), and Ultra-Reliable Low-Latency Communication (URLLC).

To provide the required performance for this wide range of use cases, 5G networks rely upon the concept of network slicing [6]. A network slice aims to create a virtual network on top of a common physical infrastructure. This virtual network is then customized and optimized to offer the resources, virtual network functions (VNFs), latency, and bandwidth expected to meet a particular demand [7].

In order to instantiate such virtual networks, there is a need for defining a meaningful clustering of users. Since 5G will support various types of user equipments (UEs), an understanding of the aggregated behavior of UEs is vital. The discovery of relevant groups of users who share the same characteristics and behavior means that they would be sharing similar requirements. This can reflect on the creation of network slices that can handle the demands of each users' group.

The major challenges facing such analysis are the lack of user-activity data in the first place, the large amount of data generated from devices continuously, and the analysis complexity. In addition, these constraints make pattern searching difficult from the operators' point of view.

In recent years, machine learning techniques recorded major success to deal with such limitations in other scenarios, and found optimal results in a reasonable amount of time. In this vein, this paper proposes a new solution dubbed "SOCL" for the Service-Oriented CLustering. SOCL relies on widely used clustering methods (i.e., k-means and hierarchical clustering) to discover patterns and identify relevant groups of UEs, according to their usage of the network, consumption of services, and other behavioral features. This will facilitate an optimal slice configuration, instantiation and selection, which will make the overall network more flexible, adaptive, and automated.

The remainder of this paper is organized as follows. Section II summarizes the fundamental background topics and related research works. Section III introduces the network slice planner (NSP), a simulation framework on which SOCL will be based. Section IV describes the main components of SOCL. Section V provides an analysis of the results obtained. Finally, Section VI concludes the paper with insights and future research venues.

## II. RELATED WORKS

The evolution towards 5G requires the provision of networks in an "as a service" fashion. This begins by enabling a dynamic configuration, and customization of network slices based on application-specific service level agreements (SLAs), and users' preferences.

Several management and orchestration solutions based their virtual network placement solutions on the collected resource consumption data [12], [13]. However, the intelligent automation of such process requires a prior knowledge of resource usage, and most importantly of trends and patterns in users' service consumption. Indeed, a proper classification of such data relying on machine learning techniques, will help to supervise and automate the configuration, implementation, and maintenance of network slices (i.e., automation of network slices life-cycle management).

Many attempts have been carried out to generate data on the performance and behavior of cloud networks such as CloudsimNFV [8]. However, to the best knowledge of the authors, NSP is the first to consider a user-centric simulation experience coupling both the underlying network simulation as well as the user/devices realistic behavior [9], and to provide logs that can be used by VNF placement algorithms for slice planning purposes. In this context, many works have been able to benchmark and test VNF placement algorithms [10], [11] using NSP data.

## III. NETWORK SLICE PLANNER

NSP is a simulator that mimics as much as possible the reallife service consumption relevant to 5G-related use-cases. As illustrated in Fig. 1, NSP defines a spatio-temporal modeling of mobile service usage (i.e., usage over a particular geographical area and in real time progress). The mobile services that are simulated using NSP are as follows:

- Video streaming services.
- Social network services.
- Instant messaging services.
- UAV delivery services.
- IoT sensors.

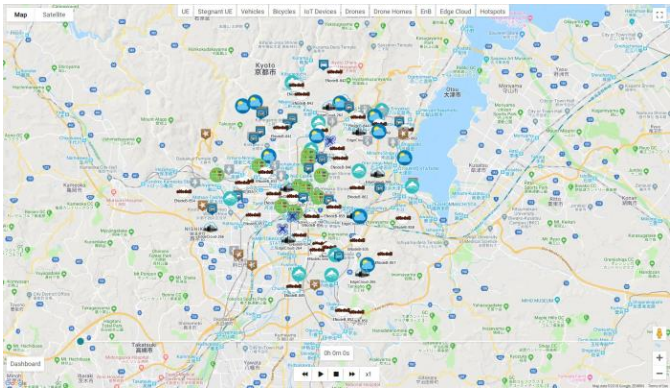

Fig. 1. NSP Simulation

Furthermore, NSP simulates the real-time edge computing resource usage in terms of RAM, CPU, storage, and record several types of events (e.g., handoff operations, trakcing area updates, attach/detach, migration, etc.). The network side is handled by leveraging NS3's LENA module, providing detailed information and statistics on Key Performance Indicators (KPIs) such as temporal variation of PHY Layer KPIs (i.e., RSRP and SINR reported by UEs, etc.) and temporal PDCP Layer KPIs (i.e., average PDU size, delay, etc.).

Within NSP, several mobility patterns are taken into account (i.e., walking, biking, or driving) using Google Direction API, which helps in the generation of real itineraries reflecting multiple users' journey.

The usage of NSP in our context is mainly motivated by the fact that users' behavioral and consumption data (i.e., mobility, service usage, network consumption, etc.) is not available. Indeed, it can be obtained either from mobile operators or directly from users' terminal, but they are reluctant to share such information, due to the critical nature and privacy constraints of such data.

## IV. SERVICE-ORIENTED CLUSTERING FRAMEWORK

Clustering in general stands for extracting natural groups of similar data objects. These groups should reflect hidden patterns in the data objects, by ensuring similarity within a cluster, and non-similarity between different clusters. In this section, we will present our NSP-based dataset, two most widely used clustering methods namely K-means (KM) and hierarchical clustering (hierarchical clustering), and their application on the dataset.

### A. NSP-based dataset

Fig. 2. NSP-based log sample

Based on the logs of NSP, the dataset that will be manipulated by the clustering methods contains the following fields before preprocessing:

- Event type: It can be an attach event, a handoff operation, or other events.
- Time stamp: Time when the data was recorded.
- Service Type: The service launched (e.g., browsing profile page, watching 720p video online, weather sensing, UAV home delivery, etc.).
- Device ID.
- Device position: The position of user or sensor in latitude longitude format.
- Edge Cloud ID: Determines the edge cloud to which the service was offloaded.
- eNB ID: Determines the eNodeB capturing the users signal.
- Track ID: Determines the tracking area to which the eNB belongs.
- Data usage: Determines the data used in KB.

Fig. 2 depicts a sample of the dataset before preprocessing. As it may be seen, the dataset contains many inputs that are less relevant to the behavior of UEs, such as event related logs. Thus, it is important to ensure an efficient preprocessing to retain only the relevant data. To be able to do so, we applied an improved implementation of the feature-selection based on mutual information and less redundancy [14], by removing before-hand all inputs that are relevant to migration, handoffs, and tracking area update, as well as the inputs of the simulation configuration.

### B. K-means

K-means is one of the most widely used unsupervised machine learning tools. Its main feature is the utilization of input vectors rather than referring to labels and obvious outcomes. This enables to discover hidden patterns. The main objective of K-means is to find a k number of clusters, each with a centroid that reflects the characteristics of the resulting group. Formally, for a given dataset $D_s = \{p_i\}$ with $i=\{1,...,n\}$ and $p_i$ is a point in the d-dimensional space $R^d$, the objective is to find a satisfying set of assignments such that the sum of squared error $\Sigma$ is minimized. $\Sigma$ is obtained using the following equation:

$$\Sigma = \sum_{j=1}^{k} \sum_{i=1}^{n} c_{i,j} \| p_i - m_j \|_2 \quad (1)$$

With $c_{i,j}$ is the cluster indicator. It equals 1 if $p_i$ and $p_j$ belong to the same cluster, and equals 0 otherwise. $m_j$ is the mean of cluster j. It is calculated as follows:

$$m_j = \frac{\sum_{p_e \in C_j} p_e}{|C_j|} \quad (2)$$

### C. Hierarchical clustering

The main motivation behind using hierarchical clustering lies in the fact that we are trying to find a hierarchical decomposition of our data. This will allow us to automate slice configuration or other operations for specific services. For instance, we can deduce a sub-cluster of one service type belonging to another service cluster. The main hierarchical clustering types are the agglomerative clustering and the divisive clustering, and both are based on a similarity matrix computed using cosine and Jaccard distance. The first family defines a bottom-up approach, meaning that the starting point is that each data point begins with its own cluster, and then a greedy strategy is followed to merge similar clusters. The second one starts by having all data points in the same cluster, and then dividing less similar clusters until the number of desired groups is reached. Within our framework, we apply the agglomerative clustering since the second type is rarely used in practice.

## V. EXPERIMENTATION AND RESULTS

In this section, we analyze the results obtained for a varying number of groups, using k-means and the hierarchical clustering methods. The main focus is on the usage of network traffic, MEC resources, and the services consumed by users in each of the obtained clusters.

### A. K-Means clustering

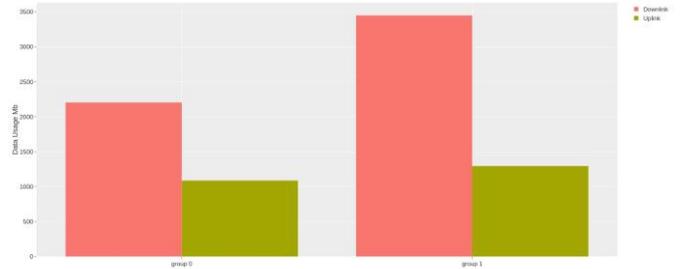

(a) Data usage

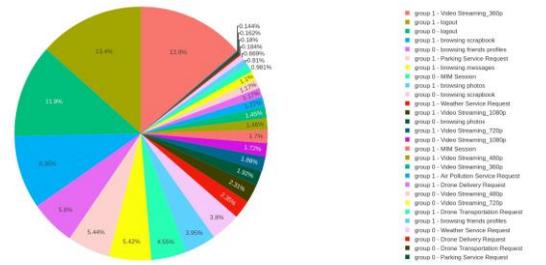

(b) Service distribution

*1) 2 clusters case:* Using K-Means with a k value of 2, users are regrouped in a logical manner where the first group consumes a low level of network traffic and MEC resources, while the second group contains users with a relatively high usage of both the network and MEC infrastructure (see Fig. 3). For the service distribution, both groups show a large usage of social networks and instant messaging services.

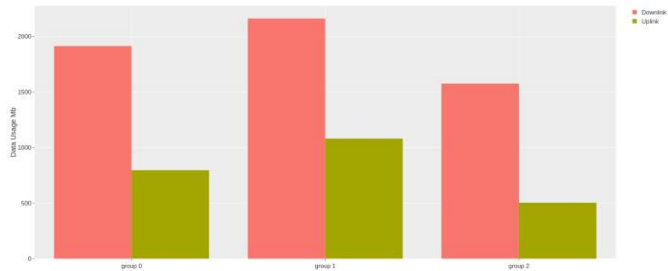

(a) Data usage

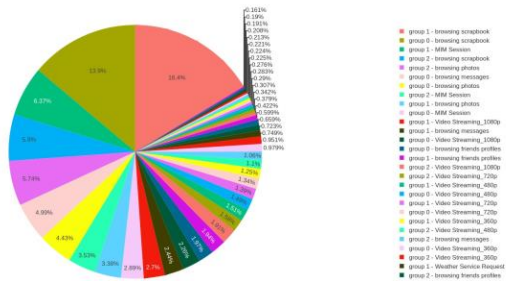

(b) Service distribution

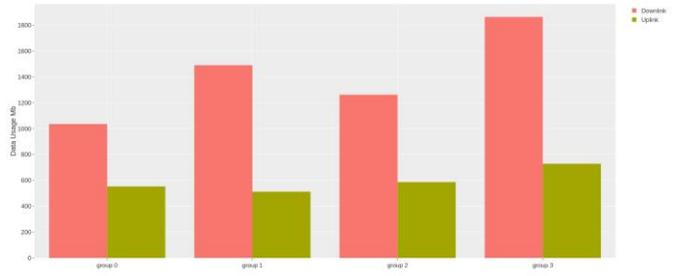

(a) Data usage

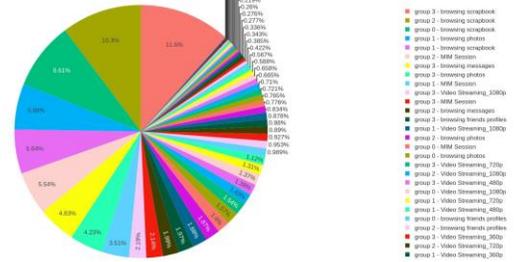

(b) Service distribution

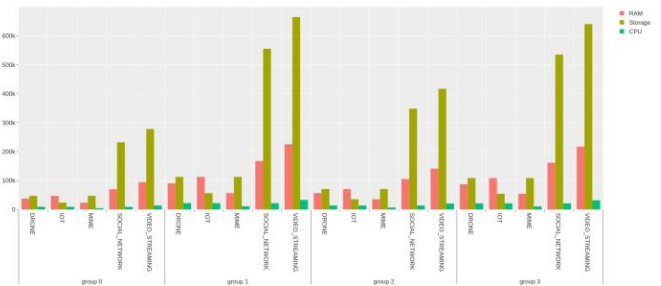

(c) Resource usage

Fig. 5. K-Means clustering with a target of 4 groups.

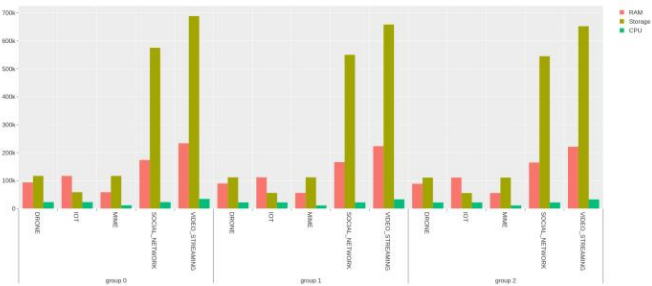

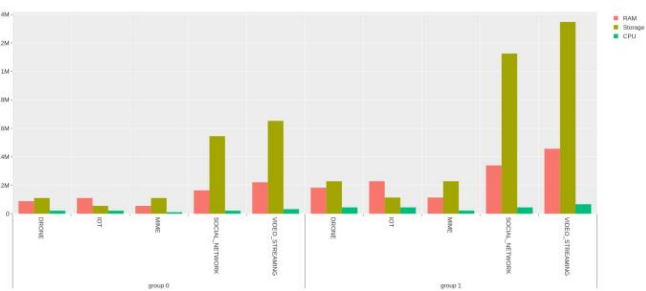

number k of clusters ( i.e., groups ), the more k-means will try to equilibrate between groups which will result in a more homogeneous set of users. In this particular case, the level of

(c)  Resource usageFig. 3. K-Means clustering with a

target of 2 groups.

(c) Resource usage

Fig. 4. K-Means clustering with a target of 3 groups.

*2) 3 clusters case:* As shown in Fig. 4, K-Means with a k value of 3 tries to balance between groups. For the characteristics related to data usage, service distribution, and resource usage, the obtained groups are equivalent with a slight difference in data usage. This equivalence is more pronounced from the resource usage perspective where the groups depict approximately the same resource usage.

*3) 4 clusters case:* Fig. 5 illustrates the scenario where kmeans is used with a k value of 4. The algorithm pursues further in the equity of groups, and the level of homogeneity within each cluster confirms that the more we increase the similarity within the obtained group can enable a prediction of virtual resources consumption, and make particular slices more available (i.e., social networks and instant messaging as the most used services).

## B. Hierarchical clustering

*1) 2 clusters case:* As illustrated in Fig. 6, when launching the Hierarchical clustering with 2 clusters as a parameter, it can be noticed that the algorithm splits the users into two groups. Although the two groups show a slightly similar behavior when comparing the data usage in the up-link, group 0 exhibits a much more usage in the down-link. For the service distribution, the two groups have a comparable service usage mostly regarding social network and video streaming services that are predominant. In terms of resource consumption, the second group (i.e., group 1 ) of users manifests a considerable overall consumption of the MEC infrastructure, especially regarding the video streaming and social network services where the consumption is nearly multiplied by two.

*2) 3 clusters case:* For the Hierarchical clustering with 3 clusters as a parameter (see Fig. 7), the first two groups display similar data usage in both the up-link and down-link, but the last group is quite different as it shows a considerably low consumption. For the service distribution, the groups have an equivalent service usage mostly of social network, instant messaging, and video streaming. The resource usage of the groups is quite ordered where the second group 0 is more resource consuming then group 1. The same remark applies to

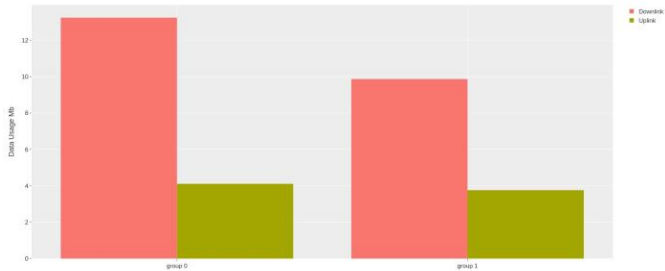

(a) Data usage

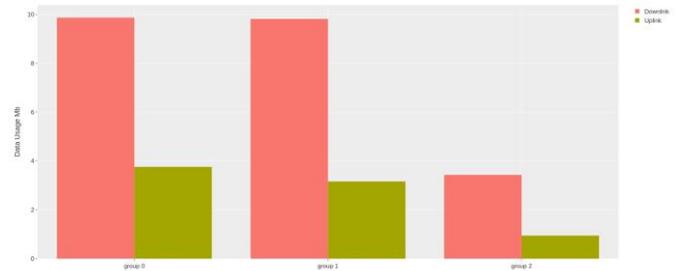

group 1 relative to group 2, with the latter having the lowest resource consumption.

*3) 4 clusters case:* As depicted in Fig. 8, the hierarchical clustering with 4 clusters regroups the users in an equitable way where we have groups ranging from a highly consuming

(a) Data usage

(b) Service distribution

(c) Resource usage Fig. 7. Hierarchical clustering with a target of 3 groups.

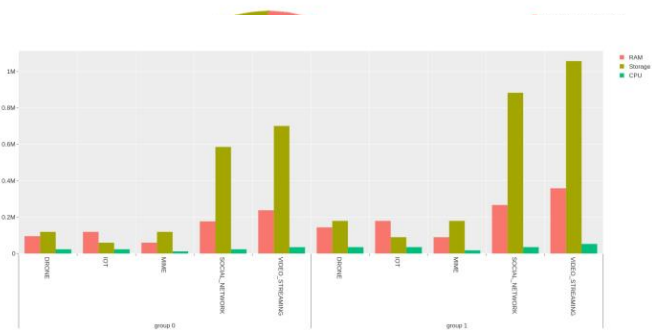

(b) Service distribution

(c) Resource usage Fig. 6. Hierarchical clustering with a target of 2 groups.

group (i.e., group 0), in terms of data usage and resource consumption, then a medium-high (i.e., group 1), a medium (i.e., group 2), and finally a group with a relatively low usage (i.e., group 3). For the service distribution, all groups have an equivalent service usage that mostly involves social network, video streaming, and instant messaging services.

## VI. CONCLUSION

Since network slicing enables the creation of logical networks tailored to support specific application demands, it is crucial to analyze the service consumption trends and the consequent resource needs. In this vein, we focus on the clustering of groups of UEs, according to their usage of the network, consumption of services, and other behavioral signs, in order to permit an optimal slice configuration, instantiation and selection. To do so, we propose SOCL as a framework for the Service Oriented CLustering, analysis and profiling of users (i.e., humans, sensors, etc.). This will be extended to enable the tailoring of the overall network for more flexibility and life-cycle management automation.

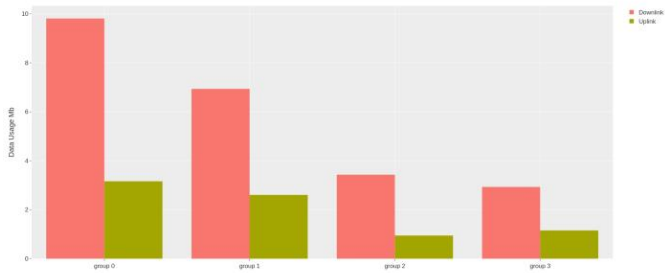

(a) Data usage

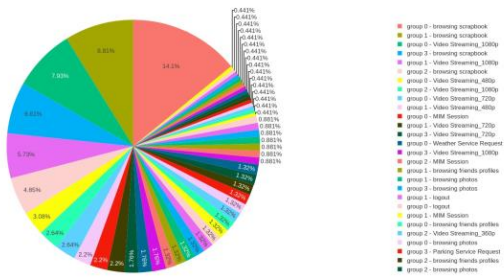

(b) Service distribution

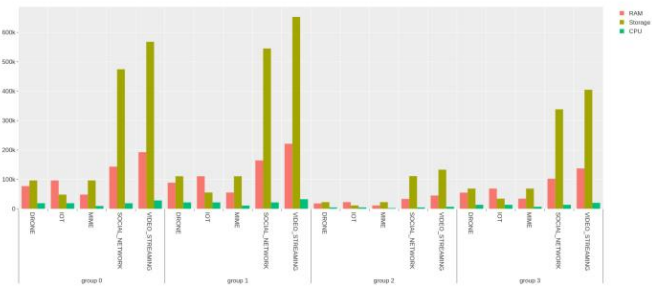

(c) Resource usage

Fig. 8. Hierarchical clustering with a target of 4 groups.

ACKNOWLEDGEMENT

This work was partially supported by the European Union's Horizon 2020 research and innovation programme under the MATILDA project with grant agreement No. 761898, and by the Academy of Finland's Flagship programme 6Genesis under grant agreement No. 318927. It was also supported in part by the Academy of Finland under CSN project with grant No. 311654.

REFERENCES

[1] 3GPP TR 23.799, "Study on Architecture for Next Generation System," rel. 14, Dec. 2016.
[2] K. Kirkpatrick, "Software-defined networking," in Communications of the ACM, vol. 56, no. 9, pp. 16-19, Sep. 2013.
[3] H. Kim and N. Feamster, "Improving network management with software defined networking," in IEEE Communications Magazine, vol. 51, no. 2, pp. 114-119, Feb. 2013.
[4] N. Paper, "Network Functions Virtualisation: An Introduction, Benefits, Enablers, Challenges & Call for Action. Issue 1," in Technical report, ETSI, Oct. 2012.
[5] M. Maternia et al., "5G PPP use cases and performance evaluation models," 5G-PPP, Apr. 2016.
[6] T. Taleb, B. Mada, M. Corici, A. Nakao and H. Flinck, "PERMIT: Network Slicing for Personalized 5G Mobile Telecommunications," in IEEE Communications Magazine, vol. 55, no. 5, pp. 88-93, 2017.
[7] I. Afolabi, T. Taleb, K. Samdanis, A. Ksentini and H. Flinck, "Network Slicing and Softwarization: A Survey on Principles, Enabling Technologies, and Solutions," in IEEE Communications Surveys & Tutorials, vol. 20, no. 3, pp. 2429-2453, 2018.
[8] W. Yang, M. Xu, G. Li, and W. Tian, "CloudSimNFV: modeling and simulation of energy-efficient NFV in cloud data centers," in CoRR, vol. abs/1509.05875, 2015.
[9] A. Laghrissi, T. Taleb, M. Bagaa, and H. Flinck, "Towards edge slicing: VNF placement algorithms for a dynamic & realistic edge cloud environment," 2017 IEEE Global Communications Conference, GLOBECOM 2017, Singapore, pp. 1-6, Dec. 2017.
[10] A. Laghrissi, T. Taleb, and M. Bagaa, "Canonical domains for Optimal Network Slice Planning," in Proc. IEEE WCNC 2018, Barcelona, Spain, Apr. 2018.
[11] A. Laghrissi, T. Taleb and M. Bagaa, "Conformal Mapping for Optimal Network Slice Planning Based on Canonical Domains, in IEEE Journal on Selected Areas in Communications, vol. 36, no. 3, pp. 519-528, Mar. 2018.
[12] T. Taleb and A. Ksentini,"Follow me cloud: interworking federated clouds and distributed mobile networks," in IEEE Network, vol. 27, no. 5, pp. 12-19, 2013.
[13] S. Clayman, E. Maini, A. Galis, A. Manzalini and N. Mazzocca, "The dynamic placement of virtual network functions," in Network Operations and Management Symposium (NOMS), pp. 1-9, 2014.
[14] H. Peng, F. Long, and C. Ding, "Feature selection based on mutual information: criteria of max-dependency, max-relevance, and minredundancy," IEEE Transactions on Pattern Analysis and Machine Intelligence, vol. 27, no. 8, pp. 1226-1238, 2005.